\documentclass[12pt,preprint]{aastex}
\usepackage{epsfig}
\usepackage{natbib}
\usepackage{emulateapj5}
\usepackage{apjfonts}
\input psfig.tex

\def\aa{{A\&A}}

\def\aj{{AJ}}
\def\al{$\alpha$}
\def\bet{$\beta$}

\def\annrev{{ARA\&A}}
\def\apj{{ApJ}}
\def\apjs{{ApJS}}
\def\asec{$^{\prime\prime}$}

\def\flux{ergs s$^{-1}$ cm$^{-2}$}

\def\hst{{\it HST}}
\def\kms{km s$^{-1}$}
\def\lamb{$\lambda$}
\def\lax{{$\mathrel{\hbox{\rlap{\hbox{\lower4pt\hbox{$\sim$}}}\hbox{$<$}}}$}}
\def\gax{{$\mathrel{\hbox{\rlap{\hbox{\lower4pt\hbox{$\sim$}}}\hbox{$>$}}}$}}
\def\simlt{\lower.5ex\hbox{$\; \buildrel < \over \sim \;$}}
\def\simgt{\lower.5ex\hbox{$\; \buildrel > \over \sim \;$}}
\def\lum{ergs s$^{-1}$}
\def\mbh{{$M_{\bullet}$}}
\def\micron{{$\mu$m}}
\def\mnras{{MNRAS}}

\def\percm2{cm$^{-2}$}
\def\peryr{yr$^{-1}$}

\def\solum{$L_\odot$}
\def\solmass{$M_\odot$}
\def\oii{[\ion{O}{2}]}

\def\mlb{$M_{\bullet}-L_{\rm{bul}}$}
\slugcomment{To Appear in {\it
The Astrophysical Journal}.}
\shorttitle{HE~0450$-$2958}
\shortauthors{KIM et al.}
\begin{document}
\title{The Host Galaxy of the Quasar HE~0450$-$2958}
\author{Minjin Kim\altaffilmark{1,2}, Luis C. Ho\altaffilmark{1}, Chien Y. 
Peng\altaffilmark{3,4}, and Myungshin Im\altaffilmark{2}}

\altaffiltext{1}{The Observatories of the Carnegie Institution of Washington,
813 Santa Barbara St., Pasadena, CA 91101.}

\altaffiltext{2}{Department of Physics and Astronomy, FPRD, Seoul National 
University, Seoul 151-742, Korea.}

\altaffiltext{3}{Space Telescope Science Institute, 3700 San
Martin Drive, Baltimore, MD 21218.}

\altaffiltext{4}{STScI Fellow.}

\begin{abstract}
A recent study suggests that the quasar HE~0450$-$2958 is hosted by a galaxy
substantially fainter than that inferred from the correlation between black
hole mass and bulge luminosity.  As this result has significant bearings on
galaxy and black hole evolution, we revisit the issue by performing an 
independent analysis of the data, using a two-dimensional image fitting 
technique. We indeed find no evidence of a host galaxy either, but, due to the 
brightness of the quasar and uncertainties in the point-spread function, the 
limits are fairly weak.  To derive an upper limit on the host galaxy
luminosity, we perform simulations to deblend the quasar from the host under
conditions similar to those actually observed.  We find that the host galaxy 
has an absolute magnitude upper limit of $-20 \lesssim M_V \lesssim -21$, 
in good agreement with the previous determination.  Since this limit is 
consistent with the value predicted from the current best estimate of the 
black hole mass, there is no compelling evidence that the quasar 
HE~0450$-$2958 has an abnormally underluminous host galaxy.  We also show 
that, contrary to previous claims, the companion galaxy to HE~0450$-$2958 
should not be be regarded as an ultraluminous infrared galaxy.
\end{abstract}


\section{Introduction}
It is widely accepted that central black holes commonly exist in massive 
galaxies and that they play an important role in their evolution (see reviews 
in Ho 2004).  This view is in large part motivated by the strong empirical 
correlations that exist between black hole mass and host galaxy properties, in 
particular the bulge luminosity (\mlb\ relation; Kormendy \& Richstone 1995; 
Magorrian et al. 1998) and bulge stellar velocity dispersion (Gebhardt et al. 
2000; Ferrarese \& Merritt 2000).   An intriguing recent observation has 
posed a challenge to this picture.  From analysis of high-angular resolution 
data on the bright, relatively nearby quasar HE~0450$-$2958, Magain et al. 
(2005; hereinafter M05) were not able to detect a host galaxy associated with 
this system.  They claim that the black hole in HE~0450$-$2958 
is hosted by an exceptionally faint galaxy, in apparent 
violation of the \mlb\ relation.  The results of M05 have stimulated lively 
debate as to the cause of this apparent anomaly, including the possibility of 
the black hole in HE~0450$-$2958 having been ejected from its companion galaxy 
in the aftermath of a merger event, either by gravitational radiation recoil 
or the slingshot effect from a three-body interaction.  Either of these
explanations, if correct, would have important implications for theories 
concerning the dynamics and evolution of binary massive black holes.

The analysis of M05, both for their {\it Hubble Space Telescope (HST)}\ images
and ground-based spectra, rely on highly specialized deconvolution methods
developed by these authors (Magain et al. 1998; Courbin et al. 2000).  While
we do not call into question the reliability of M05's analysis, given the
wide-spread interest that their results have generated it would be worthwhile
to confirm them independently, preferably using a different technique.  This
is the main purpose of this paper, where we attempt to constrain the
luminosity of the host of HE~0450$-$2958 using a two-dimensional image fitting
program.

HE~0450$-$2958 is a luminous ($M_V = -25.8$ mag)\footnote{We adopt the
following cosmological parameters: $H_0 = 100\,h = 71 $\kms~Mpc$^{-1}$,
$\Omega_{\rm m} = 0.27$, and $\Omega_{\Lambda} = 0.75$ (Spergel et al.
2003).}, radio-quiet quasar located at a redshift $z = 0.285$. Originally
discovered by de Grijp et al. (1987), it has previously been noted for being a
prominent infrared source (de Grijp et al. 1987; Low et al. 1988) with 
characteristics intermediate 
between those of starburst galaxies and quasars, prompting Canalizo \&
Stockton (2001) to speculate that the system is undergoing a transition
between these two evolutionary phases, plausibly triggered by tidal
interaction with a companion located only $\sim$1\farcs5 away.  The field
surrounding HE~0450$-$2958 is quite complex (see Fig.~1 and M05).  In addition
to the nearby, distorted companion, a bright foreground star sits 2\asec\ in
the opposite side of the quasar, and immediately adjacent to the quasar lies a
blob of ionized gas.  Given these complications and the large brightness
contrast between the quasar and any putative underlying host galaxy, we can
anticipate that any robust estimate of the host will be nontrivial (e.g.,
Schade et al.  2000).  Nevertheless, our analysis indicates that the current
limits cannot exclude the possibility that HE~0450$-$2958 has a normal host
galaxy.

\section{Data Reduction and Image Fitting}

The data analyzed in this study were retrieved from the \hst\ archive. The 
images were taken on 2004 October 1 (GO 10238; PI: Courbin) with the High 
Resolution Channel of the Advanced Camera for Surveys (ACS) through the 
F606W filter.  Three short exposures of 30~s and three long exposures of 330~s 
were taken with dithering. 
\begin{figure*}
\psfig{file=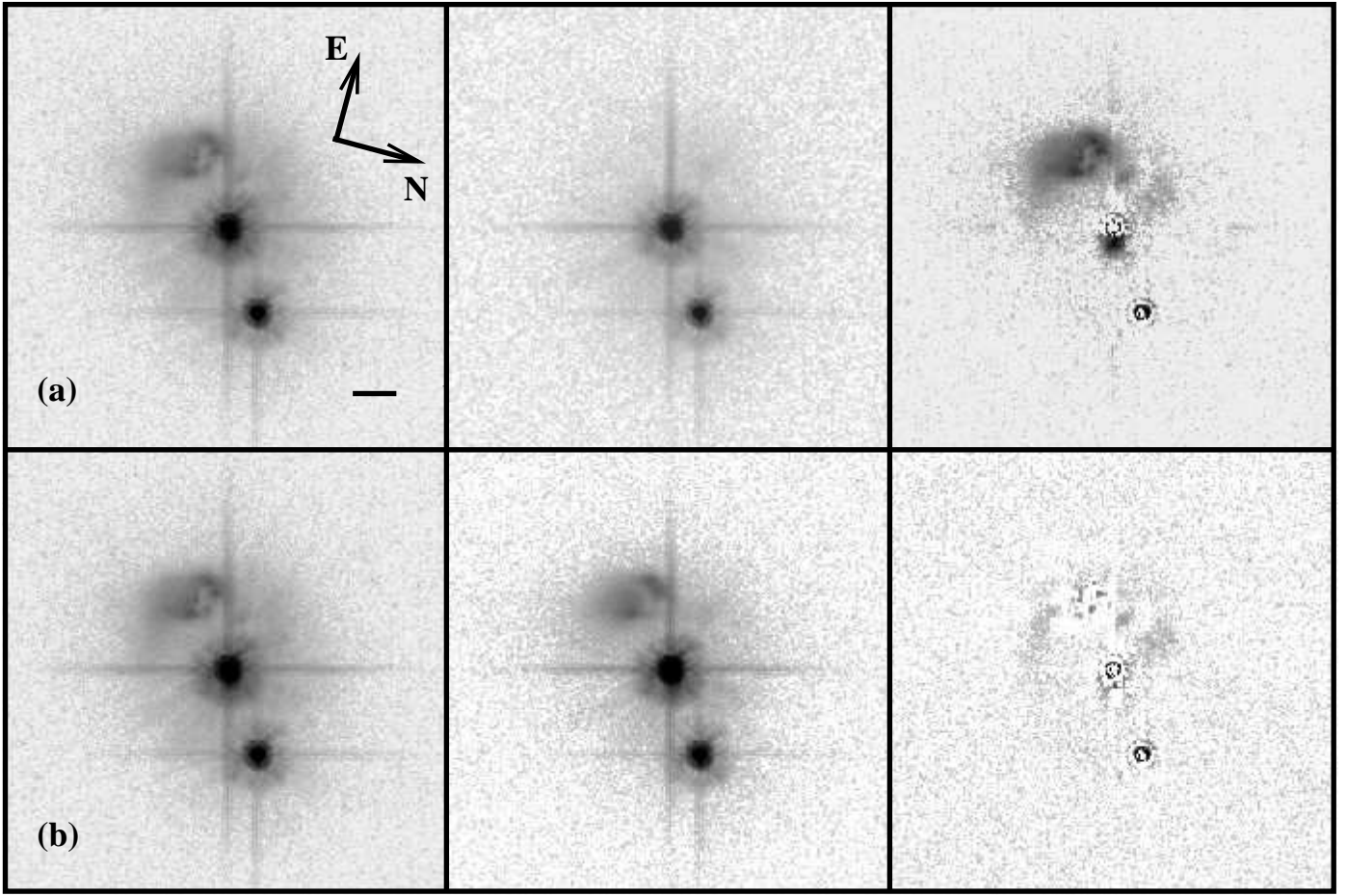,width=18.5cm,angle=0}
\figcaption[f1.eps] {
{\sc Galfit} decomposition of the {\it HST}/ACS F606W image of  
HE~0450$-$2958.  Each
panel is 10\asec$\times$10\asec, with 1\asec\ denoted by the horizontal bar.
The original, model, and residual images are shown in the left, middle, and
right columns, respectively.  In panel ({\it a}), the model contains the
quasar, a host galaxy (modeled with a S\'{e}rsic index $n=1.8$) centered on
the quasar, and the foreground star; the companion galaxy was masked out.  The
residual image clearly shows the ``blob'' adjacent to the quasar.  In panel
({\it b}), the fit includes both the companion galaxy and the blob, each
modeled with S\'ersic components of arbitrary shape described by Fourier modes.
\label{fig1}}
\end{figure*}
\begin{figure*}
\hskip 0.45in
\psfig{file=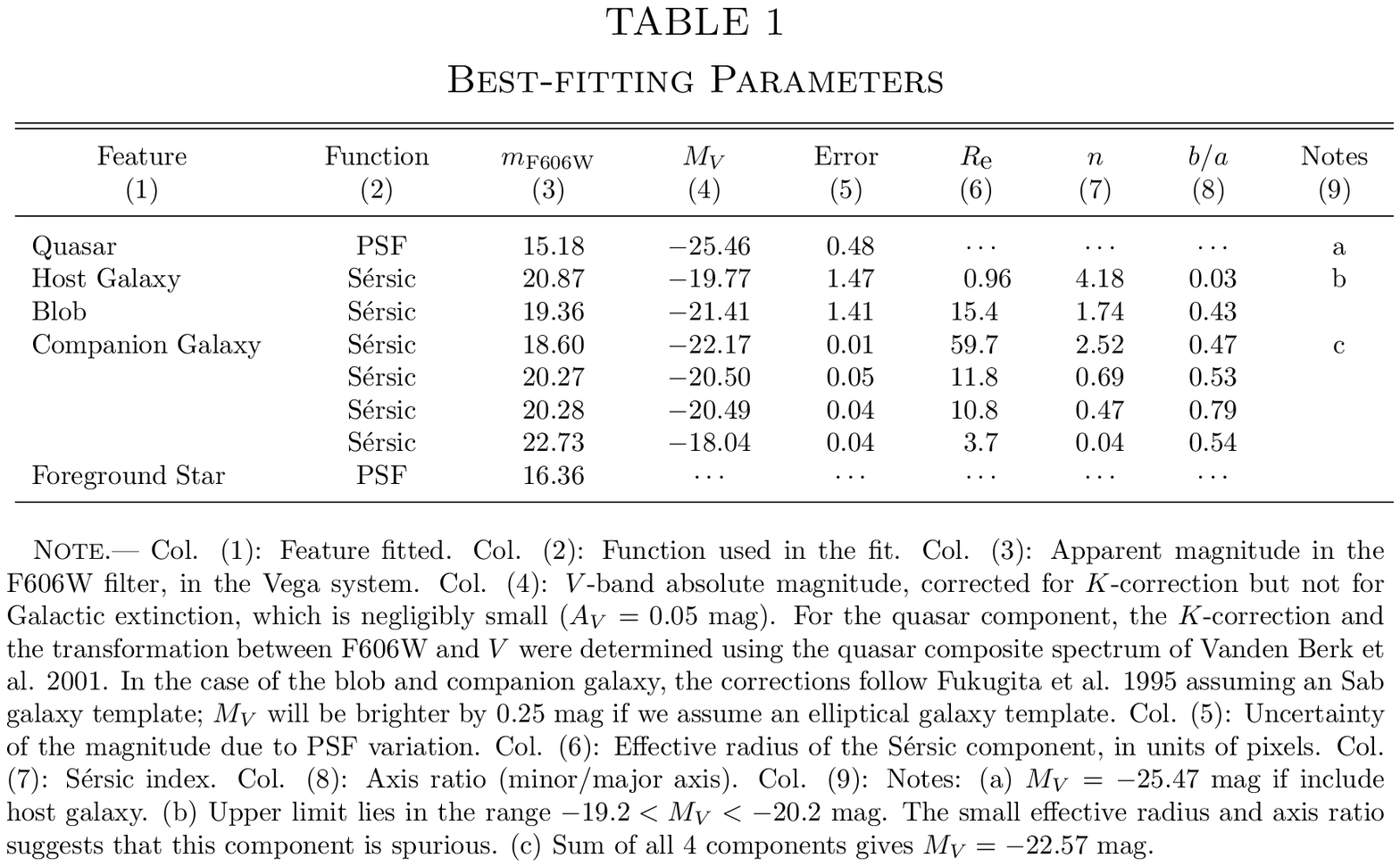,width=16.5cm,keepaspectratio=true,angle=0}
\end{figure*}
\noindent
To construct a reliable point-spread function 
(PSF), several stars, of unknown spectral type selected from the 
\textit{HST Guide Star Catalog}, were also observed with the same observing 
strategy.  One of the stars was observed during the same orbit as the 
science target, while the others were observed during different orbits; 
hereinafter we will refer to these as the ``optimal'' PSF and the
``alternative'' PSFs, respectively.
Standard data reduction steps, including bias subtraction, flat fielding, 
and flux calibration were done by the \hst\ pipeline.

We experimented with different ways to combine the drizzled subexposures, by
iteratively applying manual shifts to the PSF images.  In the end, we found
that the best results were achieved using the Pyraf-based script {\sc
Multidrizzle}, using the default coordinate information in the image headers.
After applying {\sc Multidrizzle} to each exposure subset, the saturated cores 
of the deep image were replaced by the unsaturated cores of the short-exposure 
image.

We analyzed the images using an updated version (3.0)\footnote{Documentation
and updates at {\tt
http://zwicky.as.arizona.edu/$\sim$cyp/work/galfit/galfit.html}} of {\sc
Galfit} (Peng et al. 2002; C. Y. Peng et al., in preparation), a program
designed to perform two-dimensional fits to \hst\ images of galaxies.  {\sc
Galfit} finds an optimal decomposition of a galaxy image into multiple
components, each described by a user-specified parametric model properly
convolved with an input PSF.  The choice of input PSF is critical.  For
objects such as HE~0450$-$2958, where the contrast between the active galactic 
nucleus (AGN) and the host galaxy is large, tiny mismatches of the PSF core 
will result in large mismatches in the PSF wings, and hence lead to erroneous 
inferences on the properties of the host.  To bracket the systematics induced 
by PSF variations, we perform our fits using not only the optimal PSF, but 
also the three additional alternate PSFs. The variance among the PSFs at the 
peak is $20\%$, which is consistent with the genuine PSF variance of ACS 
(Jahnke et al. 2004). Note that we do not use synthetic TinyTim (Krist 1995) 
PSFs, as these do not give a sufficiently accurate representation of the true 
PSF for the present application.

In our first attempt, we masked out the companion galaxy during the fit. We
assumed that the luminosity profile of the host galaxy can be described by a
single-component S\'{e}rsic (1968) function.  We experimented with three
cases: (1) S\'{e}rsic index $n$ = 1, which is equivalent to an exponential
profile; (2) $n$ = 4, which is equivalent to a de~Vaucouleurs (1948) profile;
and (3) $n$ was allowed to be a free parameter.  We let the program fit for
all of the other free parameters: for the host galaxy, these are its position,
luminosity, effective radius, axis ratio, and position angle; for the quasar
core and foreground star, these are their positions and luminosities.  The
case with free $n$ is shown in Figure~1{\it a}.  In the residual image, the
peak of the AGN component tends to be oversubtracted, and the blob clearly
shows up adjacent to the AGN. The best-fit component for the host galaxy has a
very small effective radius (less than one pixel).  This indicates that the
fit is unphysical, and is a symptom of {\sc Galfit} trying to assign the
residual flux from the PSF mismatch to the host galaxy component.  Varying the
input PSF did not help.

Given the proximity of the companion galaxy and the blob to the quasar, they
are likely to have a significant effect on the fit.  Thus, we next performed
the fit explicitly including these two features, using version 3.0 of {\sc
Galfit}, which models the individual components using Fourier modes to allow
for nonaxisymmetric shapes.  After some experimentation, we find that the
bright knots and the extended halo of the companion can be described by four
S\'{e}rsic components, whereas the blob can be fit with just a single
S\'{e}rsic component.  The fit using the optimal PSF is shown in Figure~1{\it
b}, and the results are summarized in Table~1.  The fit formally yields a host
galaxy with $M_V \approx -19.8 \pm 1.5$ mag, depending on the choice of PSF,
but we believe this result to be spurious, since the model has a tiny,
physically unrealistic effective radius (\lax 1 pixel) and an axis
ratio of nearly zero.  The best fit, therefore, fails to yield  
\begin{figure*}
\psfig{file=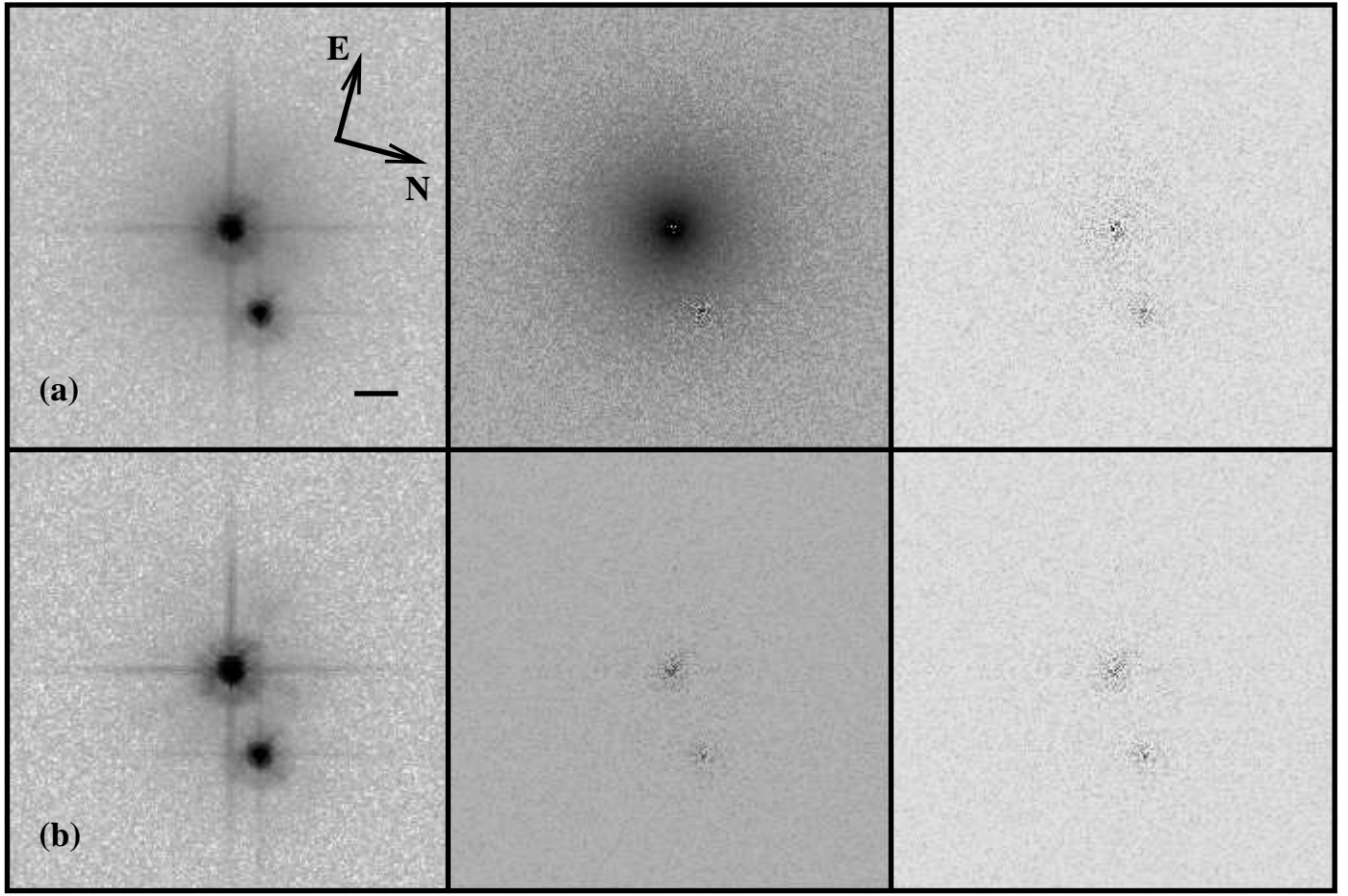,width=18.5cm,angle=0}
\figcaption[fig2.ps]
{Simulated images of HE~0450$-$2958 with an input host galaxy described by
({\it a}) $M_V=-24.2$ mag, $R_{\rm e}$ = 3\farcs4, S\'{e}rsic index $n=4$,
and $b/a = 1.0$ and
({\it b}) $M_V=-18.2$ mag, $R_{\rm e}$ = 0\farcs1, $n=4$, and $b/a=1.0$.
Each panel is 10\asec$\times$10\asec, with 1\asec\ denoted by the horizontal
bar.  The simulated image, nuclei subtracted image after the fit,
and residual image are shown in the left, middle, and right
columns, respectively.  Both the simulations and the fits were performed using
the optimal PSF.
\label{fig2}}
\end{figure*}
\vskip 0.3cm
\psfig{file=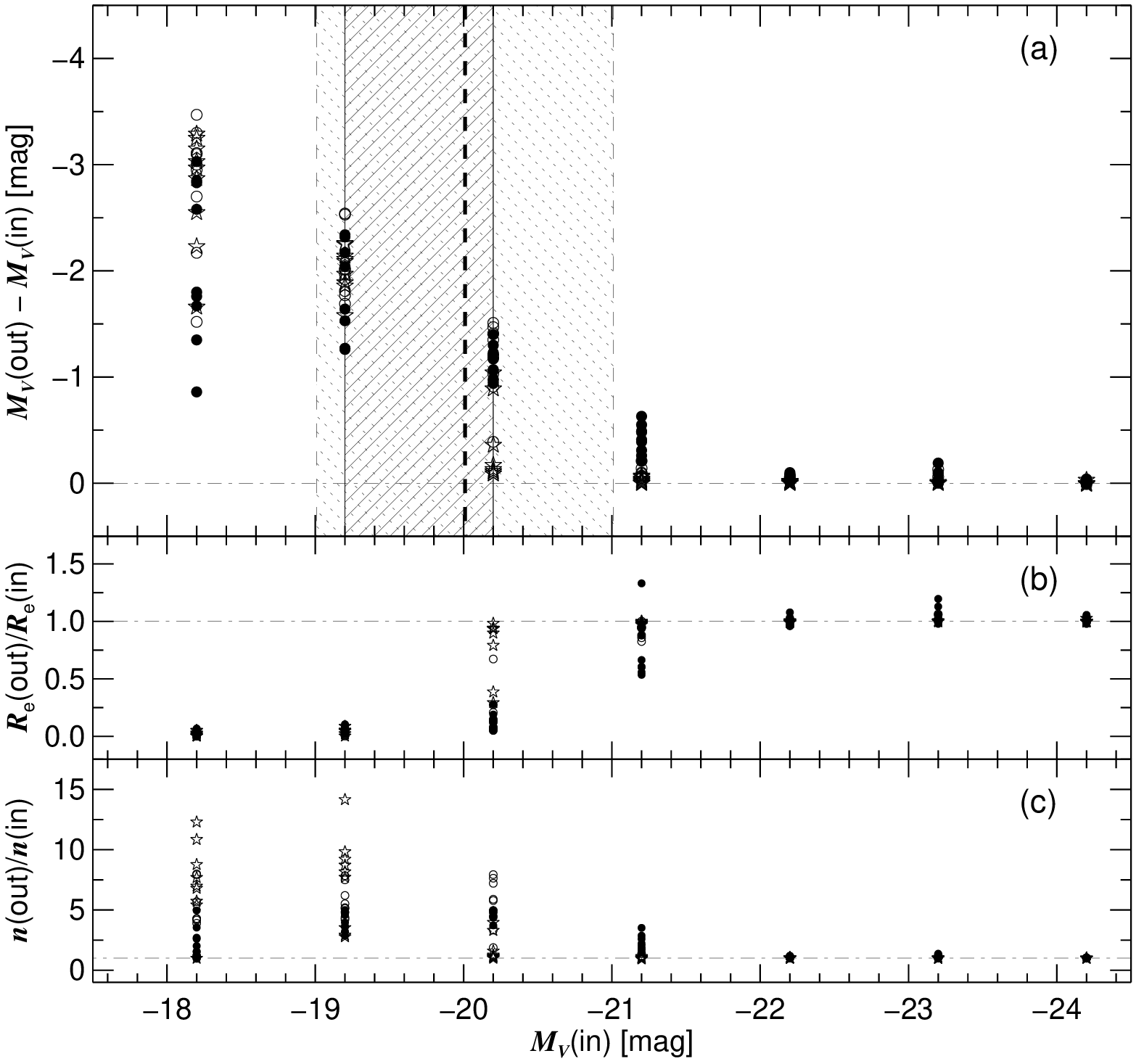,width=8.5cm,angle=0}
\figcaption[fig3.ps]
{
Simulation experiments to test the ability of our technique to recover the
properties of a hypothetical host galaxy of HE~0450$-$2958.  We show, as a
function of input absolute magnitude, $M_V$, residuals in ({\it a}) absolute
magnitude $M_V$, ({\it b}) effective radius $R_{\rm e}$, and ({\it c})
S\'{e}rsic index $n$.  We consider three models for the host galaxy: $n = 4$
({\it filled circles}), $n = 2.5$ ({\it open circles}), and $n = 1$
({\it stars}).  The effective radius of the simulated galaxies follow the
size-luminosity relation of McIntosh et al. (2005); we also show cases
corresponding to 0.5$R_{\rm e}$ and 1.5$R_{\rm e}$.  See text for details.
The dashed lines and stippled region represent the bulge luminosity (and
its 1 standard deviation) for \mbh\ = $9 \times 10^7$ \solmass\ (Merritt
et al. 2006) predicted from the \mbh-$L_{\rm bul}$ relation of Marconi \&
Hunt (2003).  The relation of  McLure \& Dunlop (2002) would predict a bulge
luminosity larger by 0.7 mag.  The solid lines and hatched region denote the
limits on the host galaxy luminosity set by our study.  The observational
limits on the host galaxy magnitude are consistent with the luminosity
expected for the revised black hole mass given by Merritt et al. (2006).
\label{fig3}}
\noindent
convincing evidence for a host galaxy.

\section{Simulations}

We first ran purely idealized simulations to determine a robust upper
limit for the luminosity of the undetected host galaxy.  Using {\sc Galfit},
we generated a set of images wherein, using the optimal PSF, the quasar and
the foreground star were placed at their respective locations.  Then,
artificial single-component host galaxies with realistic parameters were added
on top of the quasar.  Specifically, the galaxies spanned a wide range in
absolute magnitude ($-18.2 \leq M_V \leq -24.2$, $\Delta M_V = 1.0$), axis
ratio ($b/a$ = 0.7, 0.85, 1.0), and S\'{e}rsic index ($n$ = 1, 2.5, 4).  The
effective radius of each galaxy was assigned using the $r$-band
size-luminosity relation of local early-type galaxies given by McIntosh et al.
(2005), which we assume reasonably approximates the F606W band.  This
idealization represents the absolute detection limit given signal-to-noise
considerations in the absence of complications, such as PSF mismatches or
other non-random structures in the image.

We simulated a 990~s exposure by adding in an appropriate average sky
background, as given in the ACS Instrument Handbook\footnote{{\tt
http://www.stsci.edu/hst/acs/documents/handbooks/cycle15/cover.html}}, readout
noise, and Poisson noise.  Since in the real observation the bright core was
saturated in deep (990~s) image and replaced by the short-exposure (90~s) 
image, we increased the noise in the core by a factor of $\sqrt{11}$.
Finally, we use {\sc Galfit} to fit the artificial images to recover the host
galaxy, again using the optimal PSF.  Figure~2 shows two examples at the 
extremes 
\vskip 0.3cm
\psfig{file=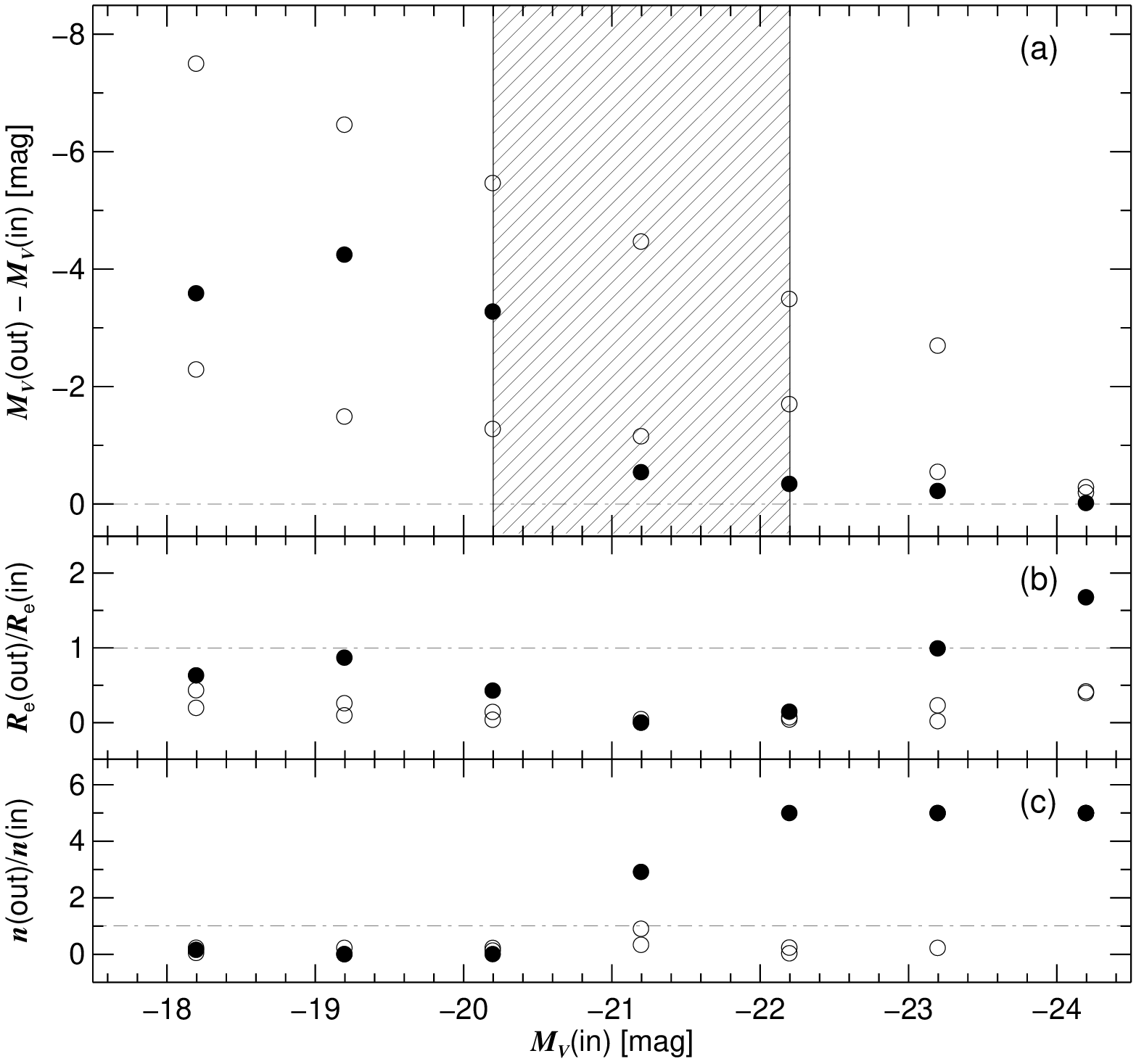,width=8.5cm,angle=0}
\figcaption[fig4.ps]
{Similar to Fig.~3, except that here we add the artificial galaxy directly
on top of the quasar in the observed image.  The galaxy is assumed to have a
S\'{e}rsic index of $n=4$ and an effective radius that follows the
size-luminosity relation of McIntosh et al. (2005).  Fits were done using the
optimal PSF ({\it solid circles}) and two of the alternate PSFs ({\it open
circles}).  In attempting to recover the host galaxy, our fits hold fixed all
the parameters, except for the amplitude,
of the quasar, blob, companion
galaxy, and foreground star, as determined from the best-fit of the original
image (Fig.~1{\it b}; Table~1).
The upper limit on the host galaxy estimated from this experiment, denoted by
the hatched region, is brighter by $\sim 1$ mag than that determined from the
more idealized situation in Fig.~3.
\label{fig4}}
\vskip 0.3cm
\noindent
of the luminosity sequence we simulated.
The results of our idealized simulations are summarized in Figure 3, where we
show, as a function of input host galaxy absolute magnitude, the output
residuals in absolute magnitude, effective radius, and S\'{e}rsic index.  As
expected, the magnitude residuals grow increasingly larger as the input galaxy
becomes fainter.  When {\sc Galfit} fails to detect the host, the S\'{e}rsic 
host component competes for the flux in the AGN core itself, resulting
in a component that is more luminous (Fig.~3{\it a}) and more compact
(Fig.~3{\it b}) than the input model.  Such a spurious object would also tend
to have a strong central peak, as evidenced by the large output S\'{e}rsic
index (Fig.~3{\it c}).

How do we establish an upper limit on the host galaxy luminosity?  As
mentioned above, when {\sc Galfit} cannot detect a host, the host component
competes for the flux from the central AGN core by shrinking in size to mimic
a PSF.  Thus, the plot for the residual of the effective radius provides an
important diagnostic.  Based on this observation, we estimate that the host
galaxy has a likely upper limit of $M_V \approx -20.2\pm1.0$ mag due to purely
signal-to-noise considerations.  Changing the input PSF to simulate PSF
mismatch does not appreciably change the qualitative behavior of the fit.

The above set of simulations, however, is highly idealized.  In order to make
the calculations more realistic, we added the artificial galaxy model directly
on top of the quasar in the {\it real}\ science image. For simplicity, we
assume an elliptical-like galaxy with $n = 4$, whose size, as above, is
specified by the size-luminosity relation. For the fit, we used the optimal
PSF plus two of the alternate PSFs.  We held fixed all the parameters for the
other components (quasar, blob, companion galaxy, and foreground star), with
the exception of their amplitude, based on the best-fitting results from the
original image (Table~1).  The trends in the residuals for this set of
simulations (Fig.~4) qualitatively resemble those for the idealized set, but
in detail they differ due to added complexities in the image.  In particular,
the optimal PSF is somewhat better than one of the alternate PSFs at
recovering the input model, thereby showing how sensitive the results are to PSF
assumptions under high-contrast imaging.

Using again the criterion of $R_{\rm e}$(out)/$R_{\rm e}$(in)$\rightarrow 0$
to gauge when the recovered galaxy parameters appear improbable, we estimate 
the upper limit of the host's absolute magnitude to be $M_V\approx-21.2\pm1.0$, 
roughly 1 magnitude brighter than in the idealized simulations.

\section{Results and Discussion}

\subsection{The Luminosity of the Host Galaxy}

This study provides a new, independent analysis of the \hst/ACS images of the 
quasar system HE~0450$-$2958.  Our analysis is based on direct two-dimensional 
fitting of the images.  Consistent with the findings of M05, we also failed to 
detect the galaxy presumably hosting the black hole.  We performed extensive 
simulations in order to place a robust upper limit on the brightness of the 
undetected host.  Depending on the assumptions adopted, our upper limit of the 
absolute magnitude of the host galaxy lies in the range $M_V\approx-20$ to 
$-21$ mag.  These limits are similar to those found by M05, based on a
very different method of analysis.

How luminous do we expect the host galaxy to be?  This can be estimated 
from the \mlb\ relation, given a black hole mass.  Assuming that the quasar is 
radiating at 50\% of its Eddington limit, M05 estimated 
\mbh\ $\approx 8 \times 10^8$ \solmass.  This value, however, is about a 
factor of 10 too large.  Merritt et al. (2006) show that HE~0450$-$2958 shares
many of the properties of narrow-line Seyfert 1 galaxies, which are thought
to be relatively low-mass black holes radiating near or perhaps even 
greater than their Eddington limit (e.g., Collin \& Kawaguchi 2004). From 
an analysis of the optical spectra taken by M05, Merritt et al. obtained a 
new estimate of the black hole mass using the virial method (e.g., Kaspi et 
al. 2000) as presented by Greene \& Ho (2005b).  Depending on the formalism 
used, they find \mbh\ $\approx (6-9) \times 10^7$ \solmass.  Choosing, for 
concreteness, \mbh\ = $7.5 \times 10^7$ \solmass, the $B$-band \mlb\ relation 
of Marconi \& Hunt (2003) for inactive galaxies predicts a bulge absolute 
magnitude of $M_B = -18.9$, or $M_V \approx -19.9$ assuming $B-V = 0.96$ mag 
for early-type galaxies (Fukugita et al.  1995).  Alternatively, if we adopt
the $R$-band \mlb\ relation of McLure \& Dunlop (2002) for active galaxies, 
properly adjusted to our cosmology, we obtain $M_R = -21.2$ mag, or 
$M_V \approx -20.6$ if $V-R = 0.61$ mag (Fukugita et al.  1995).  
Considering the allowed range of black hole masses, the current uncertainty on
the zeropoint of the virial mass estimator for AGNs ($\sim 0.5$ dex;
Nelson et al. 2004; Onken et al.  2004; Greene \& Ho 2006), and the
intrinsic scatter of the \mlb\ relation ($\sim 0.3$ dex; McLure
\& Dunlop 2002; Marconi \& Hunt 2003), the predicted host luminosity is 
consistent with the observed upper limits.

The above conclusion is subject to three caveats.  First, most nearby 
narrow-line Seyfert 1 galaxies have a disk component in addition to a bulge 
(Crenshaw et al. 2003).  Our luminosity limit for the host, therefore, is 
uncertain by its unknown bulge-to-total luminosity ratio, which for early-type 
spirals is $\sim 0.5$.  Second, we have assumed that narrow-line Seyfert 1 
galaxies obey the same black hole-bulge scaling relations as do inactive 
galaxies and other classes of AGNs.  Despite claims to the contrary (e.g., 
Wandel 2002; Mathur \& Grupe 2005), however, recent studies that directly probe 
the stellar component of the host (Barth et al. 2005; Botte et al. 2005; 
Greene \& Ho 2005a, 2006) give little reason to suspect that narrow-line 
Seyfert 1 galaxies behave abnormally.  Finally, given the quasar-like 
luminosity of HE~0450$-$2958, we should bear in mind that any direct analogy 
with the typically much less luminous narrow-line
Seyfert 1 galaxies is necessarily speculative.

\subsection{The Companion Galaxy is Not a ULIRG}

In the recent literature (e.g., M05; Haehnelt et al. 2006; Hoffman \& Loeb 
2006; Merritt et al. 2006), the companion is often referred to as an 
ultraluminous infrared galaxy (ULIRG), by which is meant a highly obscured 
system experiencing a high level of ongoing star formation.  This perception 
appears to be driven by the association of the HE~0450$-$2958 system with an 
infrared-bright source, with the implicit assumption that most of the infrared 
emission arises from stellar heating intrinsic to the companion galaxy, by the 
fact that the companion has a young stellar population (Canalizo \& Stockton 
2001; M05; Merritt et al.  2006), and the suggestion, based on its apparently 
large Balmer decrement, that the companion appears to be highly extincted 
(M05).  

We disagree with this assessment, for the following reasons.  Although the 
HE~0450$-$2958 system indeed is infrared-bright, we believe that most of the 
dust emission is associated with and heated by the quasar itself rather than 
the companion galaxy.  (The spatial resolution of the infrared observations 
is insufficient to separate the companion from the quasar.) From the 
{\it Infrared Astronomical Satellite (IRAS)}\ measurements given in 
the {\it IRAS Point Source Catalog}, the infrared flux density ratios of 
HE~0450$-$2958 ($S_{25}/S_{60} = 0.32$, $S_{60}/S_{100} = 0.81$) lie in the 
regime of AGNs (de Grijp et al. 1987; Low et al. 1988).
This indicates that the dust temperature is generally hotter than in typical 
starbursts and is more characteristic of AGN heating.  

Our second, more compelling argument is based on the serious inconsistency 
between the predicted and actual emission-line strength of the object.  The 
total (8--1000 \micron) observed infrared luminosity, $\sim 5.3\times10^{12}$
\solum\ (Canalizo \& Stockton 2001, adjusted to our adopted distance), if entirely attributed to star formation, 
would correspond to a star formation rate of 840 \solmass\ \peryr, using the 
calibration of Bell (2003). 
This estimate is severely at odds with the optical spectrum 
of the companion galaxy, which shows only weak emission lines superposed 
on a {\it post}-starburst spectrum (Canalizo \& Stockton 2001; Merritt et al. 
2006).  According to the calibration of Kewley et al. (2004), and assuming 
solar abundances, a star formation rate of 840 \solmass\ \peryr\ would 
generate a luminosity of $1.2\times10^{44}$ \lum\ for the \oii\ \lamb 3727 
emission line, or a line flux of $f_{\rm [O~II]} = 4.8\times10^{-13}$ \flux\ for
a distance of 1452 Mpc.  This value is over 800 times larger than the actual 
flux ($f_{\rm [O~II]} = 5.9\times10^{-16}$ \flux) we measured from the 
flux-calibrated spectrum published by Merritt et al. (2006), kindly sent to us 
by T.  Storchi-Bergmann.  Note that this value is probably an upper limit, 
since it is likely that there is some contamination from the narrow-line 
emission of the nearby quasar.  One cannot appeal to high levels of extinction 
to hide the \oii\ emission, because the Balmer decrement of the companion is 
actually rather modest, contrary to the assertion of M05.  In the observed 
spectrum, the H\bet\ line indeed does appear to be rather weak with respect to 
H\al, but this is partly due to dilution with strong H\bet\ absorption from A 
stars.  A proper measurement of the Balmer decrement requires careful 
decomposition of the starlight from the line emission (e.g., Ho et al. 
1997).  After subtracting a synthetic model for the starlight from the 
observed spectrum, Merritt et al. find H\al/H\bet\ = 4.6, which, for a 
standard Galactic extinction curve, corresponds to an extinction of $A_V 
\approx 1.5$ mag---certainly non-negligible, but not much larger than for
late-type spirals.  Correcting the \oii\ line for this level of extinction
would raise the inferred star formation rate by a factor of 8.8, to $\sim 8.9$
\solmass\ \peryr, but still far less than needed to match the value deduced
from the observed infrared luminosity.

From the above considerations, one cannot escape the conclusion that the 
companion galaxy to HE~0450$-$2958 is most likely {\it not}\ a ULIRG.  While 
the close proximity and distorted morphology of the companion galaxy certainly 
suggest that it is interacting with the quasar, the lack of any concrete 
evidence that it is a massive, gas-rich merger of the ULIRG variety diffuses 
some of the recent motivation for considering merger-driven scenarios to 
interpret this system.

\section{Summary}

We perform the two dimensional fit to the \hst/ACS image of the bright 
quasar HE~0450$-$2958 and several detailed simulations to investigate the 
extent to which this object lacks a surrounding host galaxy, as suggested 
in the recent literature.  The host galaxy is not detected in our 
two-dimensional fitting.  Our simulations indicate that the likely 
upper limit of the host galaxy luminosity is $-20 \lesssim M_V \lesssim
-21$ mag, depending on the assumptions adopted.  Our upper limits are very 
similar to the value of $M_V \approx -21.2$ mag determined by M05, using a 
completely independent technique based on deconvolution.  Considering the 
black hole mass for HE~0450$-$2958 revised by Merritt et al. (2006), 
\mbh\ $\approx\, (6-9) \times 10^7$ \solmass, these limits are not in conflict 
with the luminosity of the host predicted from the \mlb\ relation, provided 
that the bulge-to-disk ratio of the host is not exceptionally unusual.  There 
is no evidence that HE~0450$-$2958 contains a ``naked'' quasar or one hosted by 
an anomalously faint galaxy.

Lastly, we point out that the companion galaxy to the quasar is most likely 
not an ultraluminous infrared galaxy.  The star formation rate inferred from 
the infrared emission far exceeds the amount estimated from the strength of 
the \oii\ \lamb 3727 line.  We argue that most of the infrared 
emission is associated with the quasar itself.

\acknowledgements 
The work was supported by the Carnegie Institution of Washington and by NASA
grant HST-AR-10969.03 from the Space Telescope Science Institute (operated by 
AURA, Inc., under NASA contract NAS5-26555). M.~K. and M.~I. acknowledge the 
support from the BK21 program and grant R01-2006-00-10610-0
provided by the Basic Science Research Program of the Korea Science and 
Engineering Foundation. C.~Y.~P. acknowledges support through the STScI 
Institute Fellowship program. We thank Thaisa Storchi-Bergmann
for sending us the reduced ground-based spectra of HE~0450$-$2958.  We are 
grateful to the referee for a timely and helpful review.

\end{document}